\documentclass[conference]{formatting/IEEEtran}

\usepackage{epsfig}
\usepackage{graphics}
\usepackage{amsmath}	   \usepackage{mathrsfs}      \usepackage{shadow}        \usepackage{hhline}        \usepackage{latexsym}      \usepackage{subfig}     \usepackage[hyphens]{url}  \usepackage{cite}		   \usepackage{xspace}		   \usepackage{multirow}
\usepackage{rotating}
\usepackage{pgfplots}
\usepackage{color}       \usepackage{xcolor,colortbl}
\usepackage[super]{nth}
\usepackage{tikz}
\usetikzlibrary{automata,arrows}

\usepackage[ruled,vlined,linesnumbered]{algorithm2e}
\usepgfplotslibrary{dateplot}
\pgfplotsset{compat=1.13}

\usepackage{shortcuts}

\renewcommand{\paragraph}{\vspace{3pt}\noindent\textbf}
\begin{document}

\title{Malware Lineage in the Wild}

\author{}

\author{

\IEEEauthorblockN{Irfan Ul Haq}
\IEEEauthorblockA{IMDEA Software Institute\\
irfanul.haq@imdea.org}

\and

\IEEEauthorblockN{Sergio Chica}
\IEEEauthorblockA{IMDEA Software Institute\\
sergio.chica@imdea.org}

\and

\IEEEauthorblockN{Juan Caballero}
\IEEEauthorblockA{IMDEA Software Institute\\
juan.caballero@imdea.org}

\and
\IEEEauthorblockN{Somesh Jha}
\IEEEauthorblockA{University of Wisconsin\\
somesh.jha@cs.wisc.edu}}

\maketitle

\begin{abstract}

Malware lineage studies the evolutionary relationships 
among malware and has important applications for malware analysis.
A persistent limitation of prior malware lineage approaches is to 
consider every input sample a separate malware version. 
This is problematic since a majority of malware are packed and the packing 
process produces many polymorphic variants 
(i.e., executables with different file hash) 
of the same malware version.
Thus, many samples correspond to the same malware version and it is 
challenging to identify distinct malware versions from polymorphic variants.
This problem does not manifest in prior malware lineage approaches because 
they work on synthetic malware, malware that are not packed, or packed malware
for which unpackers are available. 

In this work, we propose a novel malware lineage approach that works on 
malware samples collected in the wild.
Given a set of malware executables from the same family, for which no
source code is available and which may be packed, our approach
produces a {\em \graph} where nodes are versions of the family and
edges describe the relationships between versions. 
To enable our malware lineage approach, we propose the first technique
to identify the versions of a malware family and a scalable
code indexing technique for determining shared functions 
between any pair of input samples.
We have evaluated the accuracy of our approach on 
$13$ open-source programs and have applied it to 
produce {\graph}s for \numfamilies popular malware families.
Our malware \graphs achieve on average a $26$ times reduction from
number of input samples to number of versions.

\end{abstract}

\section{Introduction}
\label{sec:intro}

Malware lineage studies the evolutionary relationships among malware, 
which has important security applications in the context 
of malware analysis.
For example, lineage can be a fundamental step for 
triage, labeling, categorization, threat intelligence, provenance, and 
authorship attribution.
The goal of malware lineage is to produce a {\em \graph} 
where nodes are versions of the family and
edges describe the ancestor-descendant relationships between versions.

Similar to benign programs, malware families evolve to adapt to
changing requirements by adding new functionality, and to improve
stability by fixing bugs.  However, malware development typically
comprises of an extra step not present in benign software development.
Once a new version of a malware family is ready, the malware authors
pack the resulting executable to hide its functionality and
thus bypass detection by commercial malware detectors.  The packing process
takes as input an executable and produces another executable with the
same functionality.  The packing process is typically applied many times
to the same input executable, creating polymorphic variants (i.e., executables) 
of exactly the same version, 
which look different to malware detectors. 

An important open problem in malware lineage is identifying the versions of 
a malware family among a set of input executables belonging to the family. 
The study of malware lineage goes back over $20$ years and multiple
approaches have been
proposed~\cite{sorkin1994grouping, hull1995b, goldberg1996constructing, carrera2004digital,
  karim2005malware, ma2006finding, wehner2007analyzing, gupta2009empirical, khoo2011unity, Lindorfer2012lineage, iline, darmetko2013inferring}.
However, a persistent limitation of these approaches is that they
consider every input sample a separate malware version. 
Thus, their output \graphs have a node for every input sample.
This is problematic because the packing process 
produces many polymorphic variants of a malware version. 
Thus, many input samples should be represented by the same node in the \graph.
This problem does not manifest in prior malware lineage approaches 
because they are evaluated on synthetic malware, 
malware that is not packed, or packed
malware for which unpackers are readily available (e.g.,
UPX~\cite{upx}).
But, the majority of malware is packed 
and malware often uses custom packers for which off-the-shelf unpackers 
are not available~\cite{Ugarte14}. 
Thus, version identification needs to be addressed 
with malware collected in the wild.

Identifying versions among packed samples from a malware family
is a novel and challenging problem.
This problem differs from 
program similarity~\cite{flake2004structural} 
because two versions of the
same family may be highly similar, but still different versions. 
For example, one version may patch the previous one by adding a 
conditional to fix an error condition. While the two versions may be nearly identical, 
they need to be represented by different nodes in the \graph.
We address this problem by considering malware samples with the 
same set of functions as polymorphic variants, 
which should be represented by the same node in the \graph. 
Any changes in functionality between two samples  
such as adding a function, removing a function, or 
updating an existing function, means that both samples are from different 
versions, with their own nodes in the \graph.

Another challenge with malware collected in the wild is that we do not 
know how it was developed. 
Thus, we need a lineage inference algorithm that works independently of the 
development model used by the malware authors, 
e.g., straight-line, multiple independent lines, branching and merging.
Unfortunately, \iline~\cite{iline}, 
the state-of-the-art lineage inference algorithm, 
uses separate algorithms for straight-line and 
branching and merging development. 
Thus, \iline requires knowing in advance the development model of the software 
to select the lineage algorithm. 
This is problematic with malware since the development model is not known. 

In this work, we propose a novel malware lineage approach
that works on malware samples collected in the wild.
Given a pool of malware executables from the same family, 
for which no source code is available and which may be packed
with an unknown packer,
it produces a \graph. 
Our malware lineage approach improves the state-of-art in two ways. 
First, we propose the first technique to identify the different versions 
of a malware family present in an input set of potentially packed executables.
In our \graph, a node identifies a version of the malware
family and represents all input samples that are polymorphic variants
of that version, which greatly reduces the number of nodes compared to prior 
approaches that create one node per input sample.
Second, we propose a novel lineage inference algorithm that works independently
of the development model used by the malware, and that improves the accuracy 
compared to \iline, the state-of-the-art lineage inference approach.

Performing lineage inference on malware samples collected in the wild 
requires addressing other challenges such as 
malware clustering~\cite{Bayer09,Perdisci10,bitshred,firma}, 
unpacking~\cite{omniunpack,polyunpack,renovo,ether,codisasm}, and 
disassembly~\cite{Kruegel04,Linn03}.
To address these challenges we adapt existing state-of-the-art solutions. 
In other words, our goal is not to propose novel malware clustering, 
unpacking, and disassembly techniques. 
Rather, we want to understand how far we can get using state-of-the-art 
techniques and to identify areas where improvements are needed.
To this end, we propose two novel metrics to quantify the accuracy of
the malware unpacking and disassembly process.

Our approach works as follows:
Since malware executables are collected in the wild without labels 
indicating their family, we first cluster input executables into 
malware families.
Then, each malware sample is unpacked using a generic dynamic unpacker.
The unpacker recovers the original code as raw byte sequences in 
memory snapshots.
To enable malware lineage, we need to represent the code in a form 
that enables further analysis. 
For this, we disassemble the unpacked malware code by removing code overlaps,
identifying function boundaries, and applying
dynamic information to improve disassembly results. 
This process outputs the unpacked and disassembled code as an \db~\cite{ida}, 
which serves as input to the lineage inference module that produces the 
\graph.

We have implemented our malware lineage approach and evaluated its
accuracy on open-source programs, before applying it to \numfamilies malware
families.  Our unpacking and disassembly modules recover up to 81\% of
the original functions of programs packed with $\numpackers$ packers.
The missing functions are properly unpacked, but the function
identification misses their start addresses
preventing their disassembly.  We have evaluated our lineage inference
module on \numbenignversions versions of $13$ open-source programs, 
covering over $59$ years of software development.
Our {\graph}s achieve an average 95\% accuracy, 
which improves on \iline's results, without requiring apriori 
knowledge of the malware development model.
We also show that our version identification technique has better 
accuracy and efficiency compared to using \bindiff.

We have also evaluated our approach on \nummalware packed malware samples from 
\numfamilies malware families. 
The generated {\graph}s show that our approach can handle different 
malware development models, 
i.e., straight line, $k$ independent lines, branching and merging.
The {\graph}s succinctly summarize the evolution 
of a malware family, achieving an average $26$ times reduction 
from number of input samples to number of versions.
For example, the $1,354$ \sytro samples are grouped into $6$ versions 
in the \sytro \graph.

\noindent Our contributions are:
\vspace{-6pt}
\begin{itemize}

\item We propose a novel approach for malware lineage that works with 
  malware collected in the wild.
  It takes as input a set of samples from the same malware family and 
  outputs a \graph that describes ancestor-descendant relationships among the 
  family versions. 

\item We present the first technique for identifying 
  the versions present in a pool of samples from the same malware family.  
  Our technique classifies the input samples into versions, 
  identifying polymorphic variants of the same version.

\item We propose a novel lineage inference algorithm that works independently 
  of the development process used by the malware and that improves the 
  accuracy compared to the current state-of-the-art solution.

\item We propose two metrics to quantify the accuracy of the 
  unpacking and disassembly process. 
    
\item We evaluate the accuracy of our lineage approach on 
    13 open-source programs, 
    achieving 95\% accuracy on the \graphs.   
  We apply our approach to 
    \numfamilies malware families achieving a $26$ times reduction from 
  input samples to versions.

\end{itemize}
\section{Overview \& Problem Definition}
\label{sec:overview}

Our problem is given a set of malware samples from the same 
malware family collected in the wild, 
to build a {\em \graph} that captures the evolution of the 
malware family across versions. 
Section~\ref{sec:challenges} details the challenges introduced by working on 
malware collected in the wild.
Section~\ref{sec:graph} details the problem. 
And, Section~\ref{sec:versions} describes the version 
identification subproblem.  

\subsection{Challenges}
\label{sec:challenges}

The main requirement for our malware lineage approach is 
to handle malware samples collected in the wild. 
This entails the following challenges: 

\begin{itemize}

\item{\bf Unknown versions.}
We ignore how many versions a malware family has,
i.e., we do not know apriori the nodes in the \graph.
Instead, we need to identify which input samples correspond to
different versions and which are polymorphic variants of the same version.
This is a key difference from our approach with respect to 
prior malware lineage approaches~\cite{hull1995b,goldberg1996constructing,carrera2004digital,karim2005malware,wehner2007analyzing,gupta2009empirical,khoo2011unity,iline,darmetko2013inferring} that consider each input sample as a distinct 
version.

\item{\bf Unknown development model.}
We ignore how a malware family is developed.
The current state-of-the-art malware lineage approach is \iline~\cite{iline}, 
which proposes different malware lineage algorithms for different development 
models, e.g., one for straight-line and another for branching and merging.
But, with malware we do not know apriori which development model has been used.
By contrast, we propose a single lineage inference algorithm that works 
regardless of the development model used. 

\item{\bf Unpacking.}
Malware is distributed as executables, i.e., without source code, 
and those executables are typically packed. 
Polymorphic variants are created by repacking the same version multiple times.
Our approach unpacks the input malware executables to 
recover the original code. 

\item{\bf Disassembly.}
Malware disassembly is a difficult problem~\cite{Kruegel04}.
Our disassembly module addresses code overlaps, function identification, and 
uses dynamic information to guide the disassembly. 
It outputs the unpacked program as an \db.

\item{\bf Unordered samples.}
The order in which malware samples are collected 
may not mirror the order in which their versions were developed.
Thus, our lineage inference approach does not rely on collection timestamps.

\end{itemize}

\subsection{The \GRaph}
\label{sec:graph}

Given a set of malware samples $S = \{s_1, \dots, s_n\}$ belonging to the 
same malware family, our approach outputs a \graph, 
i.e., a directed acyclic graph (DAG) $G=(V,E,F,L)$.
$V$ is the set of nodes and each node $v_i \in V$ corresponds to a 
different version of the malware family. 
Note that throughout the paper we use node and version interchangeably.
$E \subseteq V \times V$ is the set of edges, where  
an edge $ v_i \rightarrow v_j$ indicates that $v_j$ was derived from $v_i$. 
The use of a DAG enables a version to be derived from multiple parent 
versions, which may happen when two development branches merge. 

The function $F: V \rightarrow \mathscr{F}$ labels each node with  
the set of (unpacked) functions in that version, 
i.e., $F(v_i) = \{f_1, f_2, \dots, f_k\}$. 
That is, we abstract a node by the set of functions in that version 
(output by $F$). 
Two nodes cannot have the same set of functions,
i.e., $\forall i \not= j \; F(v_i) \not= F(v_j)$.
Two nodes connected by an edge share at least one function, i.e., 
$v_i \rightarrow v_j \Rightarrow |F(v_i) \cap F(v_j)| > 0$.
Note that the function sharing property is not transitive. 
Path $ v_i \rightarrow v_j \rightarrow v_k$ indicates that 
$v_i$ shares some functions with $v_j$ and 
$v_j$ shares some functions with $v_k$, 
but $v_i$ may not share any function with $v_k$. 

The function $L: V \rightarrow 2^{S}$ labels each node with the 
set of input samples that correspond to that version. 
Thus, $L$ partitions $S$ into family versions, i.e.,
$\forall i \not= j \; |L(v_i) \cap L(v_j)| = 0 \wedge \sum_{i} |L(v_i)| = n$.
The number of input samples of a version $|L(v_i)|$
intuitively captures the type of the version. 
For example, nodes representing a large number of input samples are likely to 
be major versions, 
while nodes representing a few samples may indicate beta versions where 
some new functionality is being tested. 

\begin{figure}[t]
  \centering
\begin{tikzpicture}[->,>=stealth',shorten >=0.15pt,auto,node distance=1.75cm, semithick, every node/.style={scale=1.2, text width = 2.9em, align=center, font=\small}]

    \node[state,draw=black,text=black,inner sep=0]  (P16)                     {$16{,}5$};
  \node[state,draw=black,text=black,inner sep=0]  (P367)  [right of=P16]    {$367{,}95$};
  \node[state,draw=black,text=black,inner sep=0]  (P379)  [right of=P367]   {$379{,}31$};

 \begin{scope}[every edge/.append={scale=0.9, font=\scriptsize}]
  \path (P16)     edge  node {16}  (P367)  
        (P367)    edge  node {367} (P379);
 \end{scope}

\end{tikzpicture}

  \caption{\picsys \graph built using 131 samples.
  The node label represents the number of distinct functions in the version and
  number of samples of that version. 
  The edge label is the number of shared functions.
  }
\label{fig:picsys}
\end{figure}
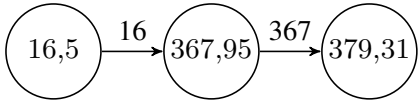

Figure~\ref{fig:picsys} shows the \graph our approach builds 
from 131 input samples from the \picsys malware family. 
Our \graph has 3 nodes, rather than 131 nodes (one per sample) 
if prior approaches were used.
The node labels are $|F(v_i)|,|L(v_i)|$, 
i.e., the number of functions in the version and 
the number of input samples of that version.
The root represents 5 input samples all with the same 16 functions.
The version following the root represents 95 samples 
that have all 16 functions in the root and an additional 351 functions. 
 
\subsection{Identifying Malware Versions}
\label{sec:versions}

Our intuition to address the version identification problem is that 
identifying versions in a pool of packed samples from the same malware family
is analogous to finding clusters of semantically equivalent executables. 
On one hand, if two executables are variants of the same version 
they should be semantically equivalent, 
i.e., provide the same functionality, 
assuming that the compilation and packing toolchains are semantics-preserving.
On the other hand, if two executables are semantically equivalent, 
then they implement exactly the same functionality and 
thus can be considered the same version. 
Note that we focus on differences visible in the binary code.
If two versions simply refactor the source code by changing
indentation or renaming variables, those changes are not visible in the
executable. Thus, their executables have the same functionality and we would 
consider them the same binary version.
That is, {\it source versions are considered different binary versions 
only if they change the program functionality}.
 
Unfortunately, current solutions to check if two executables are 
semantically equivalent are extremely expensive. 
In particular, BinHunt~\cite{binhunt} requires between 30 minutes and 1 hour to 
semantically compare two consecutive program versions. 
Our smallest malware family has 113 samples, which would require over 6 months 
to be compared by BinHunt. 
Due to the quadratic number of comparisons, 
our largest family (4,000 samples) would take over 650 years.
Other approaches perform 
{\it semantic similarity}~\cite{blex,luo2014semantics}, 
but semantic similarity is not semantic equivalence because two versions of the 
same family may have highly similar functionality, 
but still are different versions, 
e.g., a version may simply add an error condition over the previous version.

To address scalability we leverage that if two functions have the same syntax, 
then their semantics are the same (but not vice versa). 
Thus, we perform a normalized syntactic matching between executables 
to efficiently identify re-packed variants of the same version. 
For this, we first tried using syntactic similarity 
tools~\cite{bindiff,binslayer}. 
But, those tools introduce errors for version identification because, 
similar to what happens with semantic similarity approaches, 
two versions of the same family may have highly similar syntax,
but still are different versions. 
Thus, using these tools produces false positives, 
i.e., different versions identified as being the same. 
Furthermore, those tools perform pairwise comparisons, 
which create two problems for lineage inference. 
First, they do not scale well, 
e.g., after 8 hours \bindiff only completes 14\% of the pairwise comparisons 
for \winscp's 47 versions.
Second, to support branching and merging the lineage inference algorithm 
needs to identify functions that 
appear in a version, but do not appear in any predecessors
(see Section~\ref{sec:lineage}). 
Such queries cannot be performed efficiently using pairwise comparisons, 
but require an indexing approach.

To address these limitations we use normalized function hashes, 
which allow us to efficiently identify the same function 
despite instruction or block reordering and the introduction of 
padding instructions that are semantically empty. 
The function hashes can be combined into a program hash that 
identifies samples that are the same version in $O(1)$.
Function hashes also enable lineage inference to work 
because they allow us to identify in $O(1)$ all versions where a function 
is present.

\paragraph{Compilation toolchain.}
There are two cases in which we may create multiple nodes in the \graph 
for the same program version. 
First, if the same version is recompiled with different compilers or 
compilation options.
Second, if the same version is packed with different packers.
In essence, we consider the compilation and packing toolchains to 
define a version in addition to the source code.
We believe both situations, while possible, are rare with real malware. 
For example, it makes little sense to use compilation options for evasion 
(when a packer is already used for this goal) 
since recompilation does not affect all functions and instructions, and 
thus may not bypass AV signatures. 
If malware authors recompile or repack the same version with 
different toolchains our approach may overestimate the number of versions. 
Still, compared to prior work, 
our approach achieves an order of magnitude reduction
between number of input executables and number of versions in the \graph.

\section{Related Work}
\label{sec:related}

We have already compared with binary similarity approaches in 
Section~\ref{sec:versions}. 
This section describes related approaches on 
malware lineage, binary code indexing, unpacking, and disassembly.

\paragraph{Malware lineage.}
The study of malware lineage goes back over 20 years to the 
pioneering work of Sorkin~\cite{sorkin1994grouping} and the 
phylogeny of the Stoned boot sector computer virus by Hull~\cite{hull1995b}.
Most previous research on malware lineage uses a distance-based 
hierarchical clustering approach that produces a phylogenetic tree~\cite{carrera2004digital,karim2005malware,ma2006finding,wehner2007analyzing,khoo2011unity}.
Since phylogenetic trees cannot handle multiple ancestors for a node, 
other approaches such as by Goldberg et al.~\cite{goldberg1996constructing} and 
\iline~\cite{iline} use a DAG.
The above approaches analyze malware samples, 
but lineage can also analyze textual metadata 
in online threat libraries~\cite{gupta2009empirical}.

There are two fundamental differences between our approach and 
all prior work in malware lineage.
(1) We want to evaluate on real malware samples, which may be packed. 
All above approaches evaluate on unpacked malware or malware for which 
unpackers are available (e.g., UPX~\cite{upx}). 
(2) We do not know apriori the versions of a malware family,
i.e., the set of nodes in the \graph.
Instead, we classify input samples into versions and identify 
polymorphic variants of the same version.
In addition to those two differences, there are another three differences 
between our work and \iline, the current state-of-the-art for malware lineage.
(3) We assume no apriori knowledge of how the malware was 
developed and use the same approach regardless of the development model 
(straight-line, $k$ independent lines, and branching and merging).
(4) Our \graph is at a finer level of granularity.
Rather than using higher level features such as n-grams, 
individual basic blocks, or API calls observed during execution,
we focus on comparing malware samples at the level of individual
unpacked functions.
This enables identifying what functions are added and removed between
two versions and how a version is derived from its predecessors.
(5) Malware disassembly is a challenging task, which needs addressing,
e.g., \iline bypasses it by compiling the synthetic malware's source 
using {\tt gcc -S} to generate the assembly ground truth.

A related line of work addresses the problem of how to evaluate malware 
lineage approaches given the lack of ground truth. 
Hayes et al.~\cite{hayes2009evaluation} propose using artificial 
malware history generators, 
while Dumitras and Neamtiu~\cite{dumitras2011experimental} propose 
evaluating on open-source software. 
We first evaluate the accuracy of our approach on 
open-source software, before applying it to malware.

\paragraph{Binary code indexing.}
Another line of work indexes binary code to enable efficient 
code search~\cite{smit,rendezvous,expose}.
Our work also indexes the malware code to enable efficient search. 
But, those approaches do not tackle malware lineage 
and also differ in that they index call graphs~\cite{smit}, 
code fragments inside functions~\cite{rendezvous}, or 
libraries~\cite{expose}, 
rather than function and program hashes in our approach.

\paragraph{Unpacking.}
Packer identification tools~\cite{peid,rdg} use signatures to 
identify if a program is obfuscated with a specific packer. 
Those tools can be used to select a static unpacker, if available.
Generic unpackers have been proposed
to avoid manually building static unpackers for each 
packer~\cite{polyunpack,omniunpack,renovo,codisasm,rambo}.
Christodorescu et al.~\cite{Christodorescu05} and Renovo~\cite{renovo} 
propose the write-then-execute property to identify unpacking, 
used by many unpackers.
Renovo also introduces the concept of multiple 
{\em code waves} (or simply {\em waves}),
which are created by writing (i.e., unpacking) new code into memory
and then transferring execution to that unpacked code.
Code waves were later formalized by Debray et al.~\cite{Debray08tr} and 
Guizani et al.~\cite{guizani2009server}.
Guo et al.~\cite{Guo08} study the packer problem and propose heuristics 
to detect the OEP. 
Ugarte et al.~\cite{Ugarte14} propose a taxonomy of packer complexity and 
perform a longitudinal study of custom and off-the-shelf packers.
In this work we use the unpacking approach proposed by
Bonfante et al.~\cite{codisasm}.

\paragraph{Disassembly.}
Much work has addressed the challenge of disassembling binary 
code~\cite{Schwarz02,Linn03,Kruegel04,Kinder08,Kinder09,Paleari10,Zhang13,Andriesse16}, 
which in general is an undecidable problem~\cite{Wartell11}. 
Most related are approaches that address disassembly of malicious 
code~\cite{Schwarz02,Linn03,Kruegel04}.
A challenging step during disassembly is identifying the start and 
end of functions, which can be performed using a machine learning 
classifier~\cite{byteweight,Eui2015}. 
In this work we use \bw~\cite{byteweight} for function identification 
and leverage instruction and function addresses observed during 
program execution to improve the disassembly~\cite{bcr}.

\begin{figure*}[t]
\centering
\includegraphics[width=0.9\linewidth]{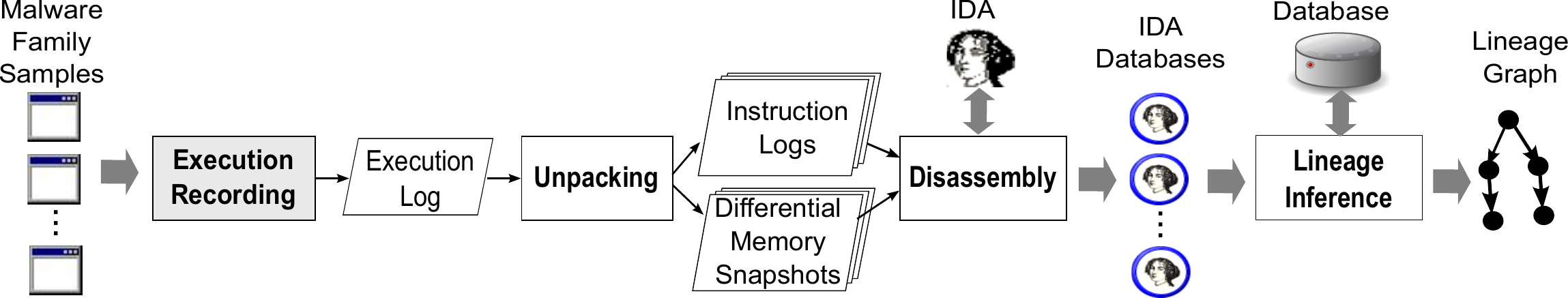}
\caption{Approach overview. The execution recording and \ida tool are 
  off-the-shelf components not developed in this work.}
\label{fig:arch}
\end{figure*}

\section{Approach Overview}
\label{sec:approach}

Figure~\ref{fig:arch} summarizes our approach. 
It takes as input a set of program executables that belong to the same 
family and outputs a \graph. 
If the input malware samples are not labeled, as is often the case, 
we first cluster them using an off-the-shelf malware clustering approach 
(not shown in the figure).
In this work, we use the AVClass tool~\cite{avclass}, 
which clusters and labels samples into known families using the AV
labels in VirusTotal~\cite{vt} reports.
While AV labels are known to be noisy~\cite{bailey2007malware,avmeter},
AVClass successfully removes noise by performing
label normalization, generic token detection, and alias detection.
AVClass outputs for each sample the most likely family name and
a confidence based on the agreement across AV engines.

Each family for which we want to analyze its lineage goes through 
processing depicted in Figure~\ref{fig:arch}. 
For each sample in the family, our processing produces 
an output \db with the unpacked version of the sample's code.
The first step in the processing of each sample is to run it 
using an {\em execution recording} module, 
which performs a lightweight recording of the execution
that can be deterministically replayed as many times as desired.
For this, we use the off-the-shelf QEMU-based 
PANDA record and replay platform~\cite{panda}.
PANDA's recording takes a snapshot of the VM state before execution and records
non-deterministic changes to the CPU state and memory during execution.
Such recording produces small changelogs 
and does not significantly slow down the VM (10\%--20\%),
which limits time-related effects (e.g. network timeouts)
on the target application.
Using a record-and-replay platform enables separating the malware execution 
from the malware analysis. 
The small logs enable efficient storage of malware executions and the 
unpacking can be rerun every time an improvement is available. 

The recording is used as input to the  
{\em unpacking}~(Section~\ref{sec:unpacking}) and
{\em disassembly}~(Section~\ref{sec:disassembly}) modules.
The unpacking module replays the recorded execution, while monitoring the 
instructions executed and the memory writes they perform.
Every time a memory area is written and then executed it 
identifies a {\em wave} and 
takes a snapshot of the memory of the program.
For each wave, the unpacking module outputs 2 {\em wave files}:
a {\em \statef} that stores the content of the memory regions overwritten 
and an {\em instruction log} with the unique instructions executed 
in the wave.
Programs with only one wave are not packed. 

The disassembly module takes as input the wave files produced by the 
unpacking and outputs an \db with the unpacked code.
It comprises of 3 steps:
{\em code region identification},
{\em function identification}, and
{\em instruction disassembly}.
For each wave, it first loads the memory ranges in the \statef into the \db,
relocating them to a different address if that range is already occupied by
previously unpacked content.
Then, it applies function identification to find the start address of 
functions present in the \statef. 
Finally, it informs \ida to disassemble inside each loaded range starting at the
instruction addresses in the instruction log, and to create functions at the 
addresses identified by the function identification module.

The \dbs for all samples in the family are input to the 
{\em lineage inference} module, 
which outputs the \graph (Section~\ref{sec:lineage}). 
The lineage inference module comprises of 3 phases. 
First, it computes the hashes of all functions in each \db and 
produces a program hash for each sample by hashing the concatenation of the 
function hashes.
Second, it builds a lineage tree 
with the most likely parent for each node. 
Finally, it adds cross-edges to the lineage tree, by identifying if any node 
has additional parent nodes from which it inherits functions. 
This last step transforms the lineage tree into a \graph.

We have implemented our malware lineage approach using over 11K 
lines of C/C++ and scripts (Python and Bash), 
as measured by CLOC~\cite{cloc},
i.e., without comments or blank lines.
Of those, 5.8K lines correspond to the PANDA unpacking plugin,
1.5K to the \ida disassembly plugin, 
3K to the lineage module, and 
1K to scripts that glue the end-to-end processing.

\section{Unpacking}
\label{sec:unpacking}

We desire four properties from our unpacking module: 
(1) Support for both off-the-shelf (e.g., Armadillo, PESpin) and 
custom packers.
(2) Support for packers of different complexity
(as defined by Ugarte et al.~\cite{Ugarte14}). 
Specifically we want to support packers for which 
the unpacked original code may not be fully available
in memory at any point of the execution, 
and samples that unpack code using multiple processes.
However, we focus on packers that do not modify the original code, 
which excludes virtualization-based packers (e.g., Themida, VMProtect).
(3) Maximize coverage, i.e., recover as much original code as possible.
(4) Minimize noise, i.e., output as little code not part 
of the original code as possible.

We use the unpacking approach proposed by 
Bonfante et al.~\cite{codisasm}, which satisfies the first three properties 
mentioned above and to which we add \statesf to support the fourth property.
Their approach runs the program and takes a memory snapshot before the 
first program instruction executes and at each wave change.
Their wave semantics define when wave changes happen and 
guarantee that any code unpacked during execution 
will be present in a memory snapshot.
Note that the unpacked code is a superset of the executed code,
i.e., the memory snapshots typically contain much ``dormant'' original code, 
which was unpacked but not executed.
For example, wave $W_i$ may unpack a function
that contains a conditional.
Only one branch of the conditional may be executed, 
but the code reachable through the other branch is also present in
the memory snapshot taken before $W_{i+1}$ starts.

There are two main differences between~\cite{codisasm} and our unpacking module. 
First, their unpacking code is not available and 
was built on top of Pin~\cite{pin}. 
In contrast, our unpacking module is implemented as a plugin for the
PANDA record and replay platform~\cite{panda},
which enables us to separate malware execution recording 
(which can be performed on dedicated malware farms~\cite{potemkin,gq}) 
from the unpacking. 
It also provides better malware isolation and allows 
the future support of other QEMU-supported architectures. 
Second, our unpacking minimizes the noise by producing a \statef for each wave, 
in contrast with the process-level snapshot produced by Bonfante et al.
The motivation for this is that a wave's snapshot shares much content
with the snapshot of the prior wave.
For example, the execution may first unpack a function $f_1$ and
execute it (first wave), then unpack another function $f_2$ and execute it
(second wave).
In this case, the first snapshot contains $f_1$,
while the second snapshot contains both $f_1$ and $f_2$.
Removing the redundancy between snapshots reduces the noise in the 
\ida database output by the unpacking and disassembly modules.

\paragraph{Unpacking overview.}
The unpacking module takes as input the execution log that records a run of 
a potentially packed program $P$. 
It replays the execution log and uses the wave semantics to identify 
waves in the list of tracked processes, 
which is initialized with the input process $P$ and to which any new 
process created (or injected from) a tracked process is added.
At the beginning of the execution of each tracked process, 
a full memory snapshot of the process address space is taken. 
New waves are detected by monitoring,  
for each instruction executed by a tracked process, 
the memory ranges it may overwrite and whether 
previously overwritten bytes are being executed. 
For each new wave of a tracked process, the unpacking outputs a \statef and a 
wave instruction log.
The \statef of wave $W_i$ stores the memory contents modified by the 
previous $W_{i-1}$ wave. 
It includes modified bytes in the main module, other modules, the heap, and 
the stack.
The wave instruction log contains 
the instructions executed during the wave and marks 
instructions following calls as function entry points.

\section{Disassembly}
\label{sec:disassembly}

The unpacking module guarantees that the code the program unpacked during 
execution is present in the \statesf. 
However, it is present as part of a long sequence of raw bytes, 
which is not particularly useful. 
The goal of the disassembly module is given as input the 
instruction logs and \statesf produced by the unpacking,
to output an \db
where the instructions and functions in the unpacked code have been identified. 
\ida is arguably the most widely used reverse-engineering tool and already 
used by most malware analysts.
Furthermore, popular IDA plugins exist for diffing two unpacked executables 
such as BinDiff~\cite{bindiff} and Diaphora~\cite{diaphora}.

Next, we detail the three disassembly steps: 
{\it memory range loading}, {\it function identification}, and 
{\it instruction disassembly}.

\paragraph{Memory range loading.}
The disassembly combines all the unpacked code into the 
same \ida database.
This includes code unpacked by all the waves observed during execution,
from the original process executed as well as 
from any child process it may create.
Intuitively, if all the code comes from the same input packed executable,
then it should be available together to the analyst
as if the input executable was not packed.
Loading the contents of the \statesf into the \db comprises of three substeps:
removing duplicate ranges, (optionally) selecting ranges, and 
relocating overlapping ranges. 
The use of \statesf prevents a memory range unpacked in wave $W_i$ 
to appear in the snapshots of waves $W_{i+1}, \dots, W_n$. 
But, some packers (e.g., YodaProtector) unpack the same content on the 
same memory range again and again. 
This behavior would introduce many duplicates of the same content in the \db. 
To prevent this, when a memory range is loaded into the \db, 
its original address and the hash of its contents are logged. 
If a range has the same original address and contents of another range already 
loaded, it is a duplicate and can be skipped.

By default, all (non-duplicate) memory ranges in the \statesf 
are loaded into the \db.
This includes ranges in the regions of the main module, other modules, 
heap, and stack. 
Some of those ranges may only contain data. 
An analyst, depending on the end application, 
may prefer to exclude data ranges to reduce the size of the final \db.
The disassembly module allows an analyst to specify 
filters on which ranges should be loaded. 
For example, an analyst could specify that ranges on the 
stack or the heap should not be loaded, or that only ranges in pages with 
execution permission should be loaded.

While loading the selected memory ranges, 
the disassembly module monitors if two memory ranges overlap.  
If so, one of them needs to be relocated because \ida 
does not support having multiple contents at the same memory range.
Internally, \ida uses a linear address space
where analysts can create segments, 
which represent contiguous chunks in the linear address space. 
Creating a segment requires the start and end addresses of the segment 
in the linear address space, and the segment base. 
The segment base is used to compute virtual addresses for the 
segments, which are the addresses an analyst sees.
The conversion from linear to virtual addresses is:
$ VirtualAddress = LinearAddress - (SegmentBase << 4)$.
The disassembly module loads each range in each wave of each tracked process 
as a separate segment. 
It starts by creating a segment at the base address of the main module of 
the initial malware process, which corresponds to the 
contents of the packed malware sample. 
For each wave of the initial malware process, 
it creates another segment following the previous one in the linear address
space\footnote{Segments are separated by a 16 byte gap}, adjusting the 
segment base so that virtual addresses in the new segment follow virtual 
addresses in the prior segment.
When a wave is relocated, the disassembly module scans 
the disassembled instructions and rewrites any addresses found in the 
disassembled code to point to the same location as before the relocation.
Once all waves of the initial process have been loaded, 
the memory range loading repeats for any other tracked processes.

\paragraph{Function identification.}
An important step when disassembling a program is identifying functions. 
Our disassembly module uses two techniques to locate function entry points and 
instructs \ida to create functions at those addresses.
First, it uses the instructions marked in the wave instruction log to have 
followed a call instruction. 
This helps identify targets of indirect calls that appeared in the execution.
Second, it uses \bw~\cite{byteweight} to locate the 
start address of functions in each of the regions loaded into the \db.
This method can identify the entry point of functions that did not execute 
during unpacking.
However, \bw identifies functions in the text section of an input executable.
Instead, we want to identify functions in the raw regions of memory
contained in the \statesf.
Thus, we have modified \bw to take as input a sequence of assembly instructions,
rather than a full executable.
Our disassembly module first disassembles 10 instructions 
starting at each offset in a region and inputs those instructions
to the modified \bw, 
which outputs a score between 0 and 1 indicating 
how likely the disassembled instructions correspond to a function start. 
As recommended in~\cite{byteweight}, the address of any sequence of 
instructions with a score higher than 0.5 is considered a function start.

\paragraph{Instruction disassembly.}
The instruction disassembly leverages the function start addresses 
identified by \bw and dynamic information from the execution, 
e.g., to identify targets of executed indirect jumps.
First, our disassembly module instructs \ida to start disassembling at 
every identified function start address. 
Then, it instructs \ida to start disassembling at every 
instruction in the wave instruction logs.
For this, the information in the instruction log is used 
to identify which wave unpacked the instruction, 
which is used to locate the \ida segment where the instruction has been loaded.
Note that when \ida is instructed to disassemble at an address, 
it performs a recursive disassembly to find as much code as possible. 
Thus, it also disassembles dormant code captured in the \statesf, 
which did not execute during the execution recording.

\section{Lineage Inference}
\label{sec:lineage}

The lineage inference module takes as input the \dbs of the family samples 
with the unpacked and disassembled code and produces the \graph.
It comprises of three phases. 
The first phase identifies the nodes in the \graph
(Section~\ref{sec:phaseI}) and also constructs the $L$ and $F$ functions
(see Section~\ref{sec:overview}). 
The other two phases identify the edges in the \graph. 
The second phase builds a lineage tree starting with 
a selected root
and greedily inserting the most similar node, 
not yet in the tree, to any node already in the tree
(Section~\ref{sec:phaseII}). 
In the lineage tree each node has at most one parent.
The third phase identifies whether some nodes inherit code from multiple 
nodes and thus need to have more than one parent
(Section~\ref{sec:phaseIII}). 
The complexity analysis of the three phases is provided in 
Appendix~\ref{sec:complexity}.

\subsection{Phase I: Identifying Versions}
\label{sec:phaseI}

Identifying the set of nodes in the \graph requires finding 
which input samples correspond to each family version.
In our approach, two input samples are the same version if they have the 
same set of functions. 
To uniquely identify a function we use a function hash. 
The use of a function hash is critical for scalability. 
It enables identifying all samples that contain a specific function in $O(1)$, 
a fundamental step in phase III.

Our approach works with different function hashes. 
Each hash can have different properties and may produce a different set
of nodes for the \graph.
Overall, we seek function hashes that determine that two 
functions are the same with low false positives, 
and are efficient to compute and index.
Currently, our approach supports two such function hashes:

\paragraph{Raw hash.}
Performs an MD5 hash of the sequence of raw byte values from function start to 
function end.  
Two functions with the same raw hash have the same bytes and thus are 
the same function. 
This hash does not require disassembly and should not produce  
false positives. 
On the other hand, it can have high false negatives when modifications 
have been applied to the function that do not affect its functionality, 
e.g., semantics-preserving instruction reordering. 
We use this hash as baseline for comparison.

\paragraph{\spp hash.}
The small prime product (SPP) 
assigns to each instruction mnemonic a small prime number. 
The hash corresponds to the product of the primes of all mnemonics 
of the instructions in the disassembled function~\cite{dullien2005graph}. 
We have modified the SPP hash computation to ignore instructions used for 
padding that do not affect the function semantics, 
e.g., the no-op (NOP) instruction and a move from 
a register to the same register.
The advantage of \spp over the raw hash is that it can detect the same 
function despite instruction reordering, block reordering, and 
instruction padding.

This phase first iterates on all the functions in each input \db, 
computing both the raw and \spp function hashes, and 
storing them in a central database.
Short functions with at most two instructions are ignored. 
Then, it computes a {\em raw program hash} and a {\em \spp program hash} 
for each sample by sorting the (raw or \spp) function hashes for the sample, 
concatenating them separated by a delimiter, and hashing the 
concatenation using MD5.
Each program hash represents one version in the \graph. 
Samples with the same program hash are polymorphic variants 
of the same version.

The raw and \spp hashes may produce different sets 
of nodes for the \graph.
The raw hash gives an upper bound on the number of nodes
since it should not have false positives.
Thus, the number of nodes using the \spp hash is always less than or 
equal to the number of nodes using the raw hash. 
While they can produce different node sets, 
both hashes often agree. 
Specifically, in 10 out of 13 open-source programs and in 
1 out of \numfamilies malware families evaluated, 
both hashes produce the same set of nodes. 

\begin{algorithm}[t]
\footnotesize
\SetAlgoLined
\DontPrintSemicolon
\KwIn{Family versions {$V = \{v_1, v_2, \dots, v_n\}$}}
\KwOut{Lineage tree T}
\Begin{
  T $\leftarrow (\{\},\{\})$\;
  // Pick smallest node as root\;
  v $\leftarrow$ minNode(V)\;
  T.addNode(v)\;
  V.remove(v)\;

  // Greedily insert nodes\;
  \While{$V \not= \{\}$}{
    // $p \in T.V, v \in V$\;
    (p,v) $\leftarrow$ findMostSimilarPair(T,V)\;
    T.addEge(p,v)\;
    V.remove(v)\;
  }
  return(T)\;
}
\caption{\Graph inference phase II.}
\label{alg:phase2}
\end{algorithm}

\subsection{Phase II: Building a Lineage Tree}
\label{sec:phaseII}

Given the nodes identified in Phase I, 
Phase II greedily builds a lineage tree where each node has a single parent. 
Its processing is described in Algorithm~\ref{alg:phase2}.
Starting with an empty tree, it first inserts as root the 
node that minimizes the sum of its size (i.e., number of functions) and 
the average distance to all other nodes.
This root selection is inspired by
Lehman's \nth{6} software evolution law (``Continuing growth''),
which states that programs tend to grow over time~\cite{lehman2001rules}.
Then, it iterates inserting at each step the node, not yet in the tree, 
with the highest number of shared functions to any node 
already in the tree.
The iteration terminates when all nodes have been inserted.

The function {\it findMostSimilarPair} greedily selects the next edge 
to insert and needs to handle two classes of ties. 
First, there could be multiple candidate nodes not yet in the tree that 
share the same number of functions with a node already in the tree. 
Here, the candidate with highest number of instructions shared with a 
node in the tree is selected.
Second, the node to be inserted could share the same number of 
functions (or instructions) with multiple nodes 
already in the tree. 
Among those, the node latest inserted in the tree 
is picked as parent.

An special case happens when the remaining nodes, not yet in the tree, 
share very few (or no) functions with nodes already in the tree
(we use an experimentally determined threshold of less than 2\% similarity). 
This may indicate independent development lines or a 
large refactoring of the code. 
In this case, the algorithm picks the smallest node not yet in the tree 
as the next to insert, rather than the one most similar to a node in the tree
(since none remaining are similar to those in the tree). 
An edge is added between the selected node and the most similar node in the 
tree. 
If the number of shared functions is zero, phase III will later remove 
those edges, thus introducing multiple independent lines.

\begin{algorithm}[t]
\footnotesize
\SetAlgoLined
\DontPrintSemicolon
\KwIn{Lineage tree T}
\KwOut{\Graph $G$}
  
\Begin{
  G $\leftarrow$ T.removeZeroSimilarityEdges()\;
  V $\leftarrow$ topologicalSort(G.V)\;
  \ForEach{$v \in V$} {
    // Find added functions (not in parent)\;
    p $\leftarrow$ G.findParent(v)\;
    $F_a \leftarrow$ F(v) $\setminus$ F(p)\;
    // Find candidate parents for cross-edge\;
    C = G.V $\setminus$ (G.successors(v) $\cup$ G.predecessors(v))\;
    // Find cross-edges\;
    X = findCrossEdges($F_a$,C)\;
    // Add cross-edges to graph\;
    \ForEach{$(p,c) \in X$} {
      G.addEdge(p,c)\;
    }
  }
  return(G)\;
}
\caption{\Graph inference phase III.}
\label{alg:phase3}
\end{algorithm}

\subsection{Phase III: Adding Cross-Edges}
\label{sec:phaseIII}

Each node in the lineage tree has at most one parent. 
However, it is possible for a version to descend 
from multiple versions, 
e.g., when a development branch is merged back into the trunk.
Phase III identifies versions with multiple parents. 

Algorithm~\ref{alg:phase3} describes this phase.
The first step is to remove any edges with zero shared functions. 
This removal introduces additional roots, identifying independent 
development lines.
Then, it iterates on the nodes in topological order. 
For the current node, it first computes the set of {\it added functions}, 
i.e., functions in this version that are not inherited from the parent version.
Then, it selects the candidate parents for a cross-edge, 
which are the nodes that are not successors or predecessors of 
the current node. 
Successors are ignored because a cross-edge from them would introduce 
a cycle in the graph. 
Predecessors are ignored because they already influence the current node.
Function {\it findCrossEdges} picks among the parent candidates the one that 
shares the most added functions with the current node. 
Then it removes those functions from the set of added functions and 
picks the next node that shares the most remaining added functions. 
The process terminates when no candidates share more than a threshold $t$ 
of functions, which we have determined experimentally to be 3.
Once {\it findCrossEdges} returns, the cross-edges are added to the graph.
Note that the use of a function hash is fundamental in {\it findCrossEdges} 
because it allows us to query in $O(1)$ time if a function in the set of added 
functions is present in any candidate parent. 

\section{Evaluation on Benign Programs}
\label{sec:accuracy}

In this section we evaluate the accuracy of our approach on benign 
open-source programs for which we have ground truth. 
Section~\ref{sec:metrics} presents the accuracy metrics, 
Section~\ref{sec:accuracy:unpacking} the unpacking and disassembly results, and
Section~\ref{sec:accuracy:lineage} the lineage inference results. 
The evaluation on malware is in Section~\ref{sec:eval:malware}.

\begin{table*}[t]
\centering
\small
\begin{tabular}{|l|l|r|rrr|rrr|}
\hline
\multicolumn{1}{|c|}{\multirow{2}{*}{\bf Packer}}  &
\multicolumn{1}{c|}{\multirow{2}{*}{\bf Version}} &
\multicolumn{1}{c|}{\multirow{2}{*}{\bf $|P|$}} & \multicolumn{3}{c|}{\textbf{FC}} & \multicolumn{3}{c|}{\bf FNR} \\
\cline{4-9}
& & & {\bf Min} & {\bf Avg} & {\bf Max} & {\bf Min} & {\bf Avg} & {\bf Max} \\ \hline
ACProtect     & 2.0      & 21 & 48.68\% & 68.73\% & 79.79\% & 23.74\% & 39.57\% & 76.00\% \\ \hline
Armadillo     & 8.20     &  2 & 61.22\% & 61.74\% & 62.26\% & 32.65\% & 35.71\% & 38.78\% \\ \hline
ASPack        & 2.12     & 21 & 46.91\% & 70.48\% & 81.53\% & 16.62\% & 24.18\% & 40.35\% \\ \hline
eXPressor     & 1.8.0.1  & 21 & 47.01\% & 70.58\% & 81.53\% & 16.74\% & 24.79\% & 43.33\% \\ \hline
FSG           & 2.0      & 21 & 46.91\% & 70.48\% & 81.53\% & 15.69\% & 21.49\% & 30.73\% \\ \hline
MEW 11 SE     & 1.2      & 21 & 46.91\% & 70.48\% & 81.53\% & 16.47\% & 22.60\% & 31.14\% \\ \hline
MoleBox       & 2.5.13   & 21 & 48.78\% & 68.63\% & 79.79\% & 27.87\% & 46.28\% & 86.55\% \\ \hline
MPRESS        & 2.19     & 21 & 49.14\% & 69.46\% & 80.14\% & 15.23\% & 22.24\% & 33.33\% \\ \hline
Packman       & 1.0      & 21 & 46.91\% & 70.48\% & 81.53\% & 15.84\% & 21.77\% & 30.87\% \\ \hline
PECompact     & 1.71     & 21 & 46.91\% & 70.48\% & 81.53\% & 16.08\% & 22.58\% & 33.33\% \\ \hline
PELock        & 2.04     & 21 & 48.78\% & 69.25\% & 79.79\% & 25.81\% & 36.61\% & 47.73\% \\ \hline
PESpin        & 1.33     & 21 & 48.68\% & 67.49\% & 79.44\% & 20.26\% & 29.25\% & 55.74\% \\ \hline
Petite        & 2.4      & 21 & 47.01\% & 70.48\% & 81.53\% & 15.97\% & 22.28\% & 32.00\% \\ \hline
RLPack        & 1.21     & 21 & 48.62\% & 69.10\% & 79.79\% & 17.02\% & 24.55\% & 39.29\% \\ \hline
UPX           & 3.91     & 21 & 47.01\% & 70.48\% & 81.53\% & 15.66\% & 21.16\% & 30.73\% \\ \hline
WinUPack      & 0.39     & 21 & 46.91\% & 70.48\% & 81.53\% & 15.69\% & 21.39\% & 30.73\% \\ \hline
YodaCrypter   & 1.3      & 20 & 48.68\% & 69.14\% & 79.79\% & 16.79\% & 24.31\% & 41.18\% \\ \hline
YodaProtector & 1.03     & 21 & 48.68\% & 68.74\% & 79.79\% & 22.37\% & 34.69\% & 71.96\% \\ \hline
\hline
\multicolumn{2}{|l|}{All} & 21 & 46.91\% & 69.26\% & 81.53\% & 15.23\% & 27.53\% & 86.55\% \\ \hline

\end{tabular}
\caption{Unpacking and disassembly accuracy evaluation on 
all SPEC CPU 2006 C programs with \spp, each packed with \numpackers packers.
For each packer it shows
the number of programs evaluated, 
function coverage, and
function noise ratio.
}
\label{tbl:accuracy_nop}
\end{table*}

\subsection{Accuracy Metrics}
\label{sec:metrics}

No prior unpacking approach proposes metrics to 
scientifically quantify the quality of the unpacking. 
Quoting a recent unpacking work:
``we recognize that we have not been able to define a metric that allow us 
to adequately determine code coverage''~\cite{codisasm}.
We propose to evaluate the accuracy of 
the unpacking and disassembly process by measuring how 
similar the original program (before packing) is to its 
unpacked representation. 
For this, we compare the \db of the original program,
disassembled using symbol information to prevent errors,
with the \db output after packing the program, and processing the packed
program with the unpacking and disassembly modules.

We propose two new metrics that
separately quantify the amount of original code recovered and
the amount of noise.
We denote the set of functions in the original program database $F_o$
and in the unpacked program database $F_u$.
We denote the set of correctly unpacked original functions
as $F^o_u \subseteq F_u$.
Correctly unpacked functions are the unpacked functions
also present in the original code, i.e., $F^o_u = F_u \cap F_o$.
Intuitively, unpacked functions that are not original code can be
considered noise, i.e., $F^n_u = F_u \setminus F^o_u$.
This noise includes, among others, unpacking functions and
falsely identified functions during disassembly.
Given the original and noise functions in the unpacked output,
we define two metrics to evaluate accuracy:
{\em function coverage} (FC) and {\em function noise ratio} (FNR).

\paragraph{Function coverage.}
The fraction of original functions the 
unpacking recovers over the number of functions in the original program. 
Ranges from zero (no original functions recovered) to one 
(all original functions recovered):
\begin{equation}
FC = \frac{|F^o_u|}{|F_o|} = \frac{|F_u \cap F_o|}{|F_o|}
\end{equation}

\paragraph{Function noise ratio.}
The fraction of noise functions over  
the total number of functions in the unpacked representation of the program. 
Ranges from zero (no noise) to one (all noise, no original functions recovered):

\begin{equation}
FNR = \frac{|F^n_u|}{|F_u|} = \frac{|F_u \setminus F^o_u|}{|F_u|}
\end{equation}

When computing FC and FNR we identify a function by its \spp hash,
and we ignore short functions,
i.e., those with at most two instructions,
since those can cause spurious matches.

\subsection{Unpacking \& Disassembly Accuracy}
\label{sec:accuracy:unpacking}

We evaluate the accuracy of the unpacking and disassembly 
modules using a dataset of 21 programs for which we have the source code: 
19 C programs from the SPEC CPU 2006 benchmark~\cite{speccpu} and 
2 C++ programs from the Olden benchmark~\cite{olden}. 
We compile them using Visual Studio with optimization level -O2 and 
debugging symbols, producing an executable and a PDB symbols file. 
We use \ida to disassemble them providing the PDB symbols file as input.
Next, we pack the executables with \numpackers off-the-shelf packers and 
examine that they work.
We find that the SPEC CPU 2006 programs do not work after packing with 
Armadillo, but the 2 Olden benchmark programs work.
There is also one SPEC CPU 2006 program that does not work after being 
packed with YodaCrypter.
We input the correctly packed executables to the unpacking and disassembly 
modules, which output an \db with the recovered code.

Table~\ref{tbl:accuracy_nop} summarizes the accuracy results
using the metrics described in Section~\ref{sec:metrics}. 
Overall, our unpacking and disassembly modules achieve an average 
function coverage of 69\% and an average function noise ratio of 27\%.
Manual analysis of the results shows that: 
(a) all original functions are present in the \statesf 
and thus in the unpacked \db, 
(b) the same functions are missed in a program regardless of the packer, 
(c) the disassembly process fails to identify 
always the start of the same functions, and 
(d) the vast majority of additional functions (i.e., noise) are added by the 
packing process to perform the unpacking at runtime.
{\it Thus, all the original code is present in the \statesf 
output by the unpacking, but, regardless of the packer, 
the disassembly process fails to identify 
the entry point of the same functions.} 
Next, we analyze this issue in more detail using the BHBmk program from 
the Olden benchmark.
We observe the same causes in other programs.

For BHBmk, the analysis reveals that \bw only finds 
37 original functions (i.e., misses 15 functions)
despite all functions being present in the \statesf 
provided as input to \bw.
Manual analysis on the missed functions shows two reasons. 
First, 9 functions are wrappers to other functions. 
These wrappers start with a 
sequence of push instructions followed by a call to the wrapped function, 
which is a common sequence in the middle of functions and thus 
not detected by \bw as a function entry point.
Second, 6 functions have uncommon preambles not in the \bw model. 
Note that even if \bw misses a function start, 
\ida may still identify the function 
if there is a direct call to it. 
Of the 15 BHBmk functions missed by \bw, \ida identifies 10 this way. 
There are also 4 functions whose start address \bw finds, 
but \ida refuses to create a function at that address.
There are two reasons for this. 
First, \ida may run into a disassembly error inside the function failing 
to find the function's boundary.
Second, \ida may have already marked that code as belonging to another 
function and refuses to create a new function there.

To conclude, the unpacking module properly outputs all original code into 
the \statesf, but not all original functions would be available to the 
lineage inference because the disassembly misses some functions.   
Most of the missed functions are due to \bw and we plan to either improve 
it or replace it in our next release. 
Most of the additional functions (i.e., noise) implement the unpacking process 
and we plan to identify those as a next step. 

\begin{table*}[t]
\centering
\footnotesize
\begin{tabular}{|l|r|r|r|r|r|r|r|r|r|r|r|r|r||r|r||r|}
\cline{4-17}
  \multicolumn{3}{c|}{} & \multicolumn{3}{c|}{\bf Reference} & \multicolumn{4}{c|}{\bf \spp} & \multicolumn{4}{c||}{\bf Raw} & \multicolumn{2}{c||}{\bf iLine (PO)} & {\bf \bindiff} \\
\hline
  \textbf{Program} & \textbf{First} & \textbf{Last} & {\bf Type} & {\bf Rel.} & {\bf $|V|$} & {\bf $|V|$} & {\bf $|R|$} & {\bf $|X|$} & {\bf PO} & {\bf $|V|$} & {\bf $|R|$} & {\bf $|X|$} & {\bf PO} & {\bf DAG} & {\bf SL} & {\bf $|V|$} \\ \hline
  FileZilla        & 3.0.0 & 3.24.0 &   S & 121 & 119 & 119 & 1 &  18 &  99\% & 121 & 1 & 20 &  97\% &  96\% & 100\% & 118 \\ \hline
Fzputtygen       & 3.0.8 & 3.24.0 &   S & 108 & \review{19} &  19 & 1 &  0 & 100\% &  30 & 1 &  0 &  96\% &  51\% & 99\% &  18 \\ \hline
Fzsftp           & 3.0.0 & 3.24.0 &   S & 118 & \review{50} &  50 & 1 &  0 &  91\% &  52 & 1 &  2 &  34\% &  38\% & 48\% &  43 \\ \hline
Notepad++        & 1.0   &  7.3   &   D &  70 &  70 &  70 & 1 &  6 &  64\% &  70 & 1 & 14 &  71\% &  76\% &  71\% &  70 \\ \hline
Pageant          & 0.50  & 0.67   &   S &  18 &  18 &  18 & 1 &  0 &  93\% &  18 & 1 &  0 &  97\% &  84\% &  99\% &  18 \\ \hline
Plink            & 0.50  & 0.67   &   S &  18 &  18 &  18 & 1 &  0 &  98\% &  18 & 1 &  1 &  90\% &  84\% &  99\% &  17 \\ \hline
ProcessHacker    & 1.0   & 2.39   & 2-S &  52 &  52 &  52 & 2 &  1 &  99\% &  52 & 2 &  8 &  79\% &  86\% & 100\% &  52 \\ \hline
PSCP             & 0.48  & 0.67   &   S &  20 &  20 &  20 & 1 &  1 &  99\% &  20 & 1 &  1 &  73\% &  96\% &  99\% &  19 \\ \hline
PSFTP            & 0.52  & 0.67   &   S &  16 &  16 &  16 & 1 &  0 &  99\% &  16 & 1 &  0 &  88\% &  95\% &  99\% &  15 \\ \hline
PuTTY            & 0.46  & 0.67   &   S &  22 &  22 &  22 & 1 &  2 & 100\% &  22 & 1 &  0 &  92\% &  98\% &  99\% &  21 \\ \hline
PuTTYgen         & 0.51  & 0.67   &   S &  17 &  16 &  16 & 1 &  0 &  99\% &  16 & 1 &  0 &  75\% &  87\% &  99\% &  16 \\ \hline
PuTTYtel         & 0.49  & 0.52   &   S &   4 &   4 &   4 & 1 &  1 & 100\% &   4 & 1 &  0 &  50\% & 100\% & 100\% &   4 \\ \hline
WinSCP           & 4.2.6 & 5.9.3  &   S &  47 &  47 &  47 & 1 &  1 &  99\% &  47 & 1 &  8 &  94\% &  99\% & 100\% &  47 \\ \hline
\end{tabular}
\caption{Lineage inference evaluation on 13 open-source programs.
For each program, it first shows the earliest and last versions analyzed. 
Then, the expected \graph type 
({\it S} for straight-line, {\it 2-S} for two straight lines, and
{\it D} for DAG), 
the number of releases, and 
the number of expected binary versions (i.e., ground truth).
Next, for the \graph generated using each hash:
the number of nodes ($|V|$) and root nodes ($|R|$),
the number of cross-edges ($|X|$), and
the partial order (PO) agreement.
The next two columns show the PO agreement using \iline's DAG and 
straight-line (SL) algorithms.
The final column shows the number of versions identified using \bindiff.
\juan{Review numbers in blue}
}
\label{tbl:versions}
\end{table*}

\begin{figure*}[t]
  \centering
  \includegraphics[width=\linewidth]{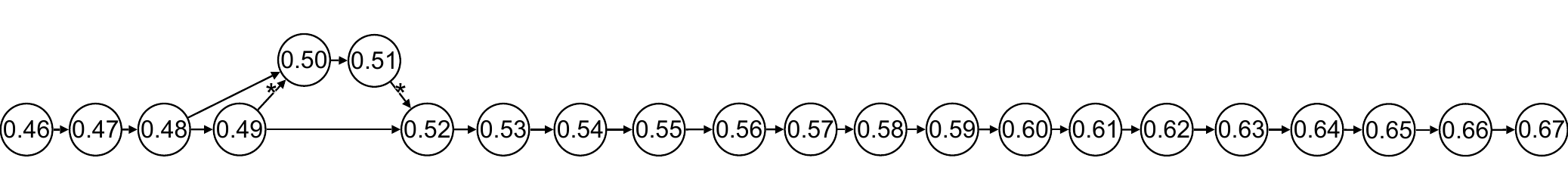}
  \caption{\putty \graph. Cross-edges are marked with asterisk.
  }
  \label{fig:putty}
\end{figure*}

\begin{figure*}[t]
  \centering
  \includegraphics[width=\linewidth]{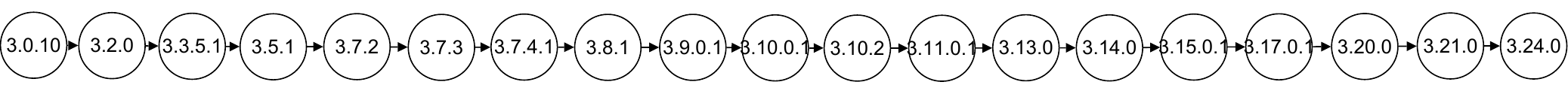}
  \caption{\fzputtygen \graph. 
  }
  \label{fig:fzputtygen}
\end{figure*}

\subsection{Lineage Inference Accuracy}
\label{sec:accuracy:lineage}

To evaluate the lineage inference we use \numbenignversions release versions 
of 13 open-source Windows programs. 
Seven of the programs are part of the \putty distribution~\cite{putty}
(\putty, \puttygen, \puttytel, \psftp, \pscp, \plink, \pageant), 
three are part of \filezilla~\cite{filezilla}
(\filezilla, \fzputtygen, \fzsftp), and 
the other three are 
\winscp~\cite{winscp},
\npp~\cite{npp}, and 
\processhacker~\cite{processhacker}. 
We are interested in evaluating the lineage inference as a stand-alone module. 
Thus, we evaluate on the released (i.e., unpacked) executables. 
To evaluate a program, we generate an \db for the executable of each version 
and input these \dbs to the lineage inference module to produce the \graph.
Then, we compare the \graph to the expected \graph built using the 
version numbers.

We manually identify the number of binary 
versions among the input releases. 
There are three main reasons why consecutive source releases 
may not create a new binary version. 
First, some versions do not modify the source code, 
but only auxiliary files (e.g., configuration files, images).
This happens, for example, between \filezilla 3.9.0 and 3.9.0.1. 
Second, some versions perform source code modifications that do not affect 
the final executable, e.g., variable renaming.
Third, in distributions with multiple programs, source code changes 
may affect some programs, but not others. 
For example, the changes between 0.61 and 0.62 
affected \putty and \puttytel, but not \puttygen.
Thus, \puttygen 0.61 and 0.62 are the same binary version. 

Table~\ref{tbl:versions} summarizes the results.
For each program it shows the first and last versions analyzed. 
Then, it shows the expected \graph type, the number of releases, and 
the number of binary versions (i.e., ground truth).
The type can be {\it S} for straight-line, {\it 2-S} for two  
straight lines, and 
{\it D} for branching and merging, i.e., DAG.
Next, it shows for the \graph generated using each hash: 
the number of nodes and root nodes, 
the number of cross-edges, and 
the partial order agreement, 
a measure of \graph accuracy proposed in \iline~\cite{iline}.
Finally, it shows the PO agreement using \iline's DAG and straight-line 
algorithms.

The results for Phase I show that the versions identified 
by the \spp hash match the expected binary versions for all 
13 programs, while the raw hash identifies the correct binary versions for  
9 programs.
For the 3 \filezilla programs, the number of nodes identified by the 
\spp and raw hashes differs, with the raw hash identifying a larger number 
of binary versions. 
In all cases, when a node in the \graph represents multiple releases 
(each with a different executable hash), 
those releases are consecutive.
These results indicate that the \spp hash is better at identifying 
versions than the raw hash. 

Analysis of the phase II results shows that 
the generated lineage tree is more accurate with the \spp hash in 9 programs and
has the same accuracy with both hashes in 4 programs. 
Thus, the \spp hash works better than the raw hash also in this phase.
The most common error is that a version $x$ is more similar to 
version $x+2$ than to version $x+1$. 
These errors happen in all but one program and manifest in two cases:
two consecutive versions swap their positions or a version has two successors 
one creating a branch with a single node and no merging point. 
An example of the first case is \puttygen where version 0.58 is 
followed by 0.60, which is followed by 0.59. 
An example of the second case is \fzsftp version 3.0.6, 
which is followed by both 3.0.7.1 and 3.0.8.1. 
The latter is followed by all other versions, while 3.0.7.1 is in a branch 
by itself.
Another error affecting 2 programs is that the wrong root is chosen, 
i.e., the chosen root is not the earliest version, but the second earliest.

All programs have the correct number of roots, 
one for straight line or DAG development, and two for \processhacker 
that has two separate lines.
For the 12 programs with straight-line (or 2-straight) development, 
phase III can only introduce errors as there is no branching and merging.
But, our algorithm should identify those cases and avoid inserting cross-edges.
The results show that cross-edges are inserted for 7 programs. 
In 5 of those programs only 1 or 2 cross-edges are inserted and these typically 
are inserted where there was an error in Phase II. 
For example, Figure~\ref{fig:putty} shows the final \graph for \putty. 
Version 0.48 is followed by versions 0.49 and 0.50 after Phase II, and 
phase III inserts cross-edges between 0.49 and 0.50 and between 
0.51 and 0.52. 
If 0.50 had followed 0.49 after Phase II, no cross-edge would have been 
inserted as 0.49 would be a predecessor of 0.50 and thus 
not included in the list of candidate parents.
The same applies to the cross-edge from 0.51 to 0.52.
Figure~\ref{fig:fzputtygen} shows the final \graph for \fzputtygen. 
In this case, the \graph has no errors and perfectly 
matches its straight-line development.

\paragraph{Comparison with \iline.}
We compare our lineage inference algorithm with \iline, 
the current state-of-the-art lineage inference approach.
\iline proposes two different lineage inference algorithms 
for straight-line code and branching and merging, 
and assumes the analyst knows which of the algorithms 
should be used (or which results to trust if both algorithms are used).
Since \iline is not publicly available we have re-implemented both of 
its algorithms.
The two rightmost columns in Table~\ref{tbl:versions} show the results for 
\iline's DAG and straight-line (SL) algorithms
using as symmetric distance the number of differing functions 
between both samples according to the \spp hash.
This allows us to compare \iline's algorithms with our algorithm 
using similar features.
To measure accuracy, we use the partial ordering agreement (PO) metric 
proposed in \iline.
For \processhacker, we first split versions into both development lines 
and apply \iline's SL algorithm in each line separately.
Our lineage inference module using \spp achieves an average 95\% PO 
over the 13 programs,
compared to 84\% for \iline's DAG and 93\% for \iline's straight line.
Thus, our algorithm outperforms both of \iline's algorithms despite 
using no apriori knowledge about the development model, 
which is fundamental when operating with malware. 
The best results for \iline are obtained by assuming straight-line development, 
but this means that a wrong (straight-line) \graph would be 
produced for any program using branching and merging.

\paragraph{Comparison with \bindiff.}
We compare our Phase I results with \bindiff. 
Pairwise comparison of all versions does not finish in 8 hours for 
\winscp and \filezilla. 
To address this, we first partition the executables by number of functions 
(i.e., executables with differing number of functions should be different 
versions) and only perform pairwise comparisons inside each partition.
After aligning two executables with \bindiff we consider they are the 
same version if \bindiff successfully aligns all the functions in both executables
and
the similarity value for each function is over 0.9.
The righmost column in Table~\ref{tbl:versions} shows that in 7 out of 13 
programs, this approach would introduce errors, 
i.e., incorrectly identifying similar versions as being the same.
Furthermore, note that we cannot use \bindiff in Phase III of our algorithm
because it cannot check if a function appears in any of the predecessor 
nodes. We need function hashes for that.

\begin{figure*}[t]
  \centering
  \includegraphics[width=\linewidth]{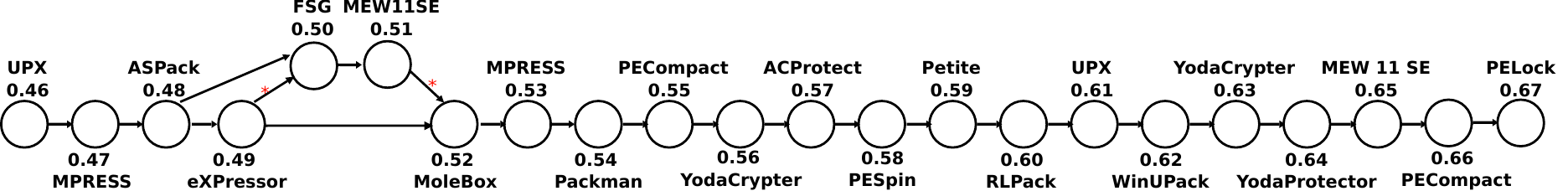}
  \caption{\Graph for \putty when each version is packed with a different packer.   }
  \label{fig:puttypacked}
\end{figure*}

\paragraph{Packed evaluation.}
To evaluate the lineage inference accuracy with packed executables, 
we perform the following experiment. 
We pack each version of \putty using a randomly selected packer and then 
apply the unpacking, disassembly, and lineage inference modules to build the 
\graph. 
This allows us to evaluate our lineage inference approach when 
different packers are used by the same program, as well as to check 
the impact on the \graph of any functions missed by the 
unpacking and disassembly modules.
Figure~\ref{fig:puttypacked} shows the produced \graph, 
which is identical to the unpacked version in Figure~\ref{fig:putty}. 

In summary, our evaluation has shown that: 
the \spp hash outperforms the raw hash both in node and edge identification;
the \graphs produced using the \spp hash are more accurate than those 
produced by \iline, without assuming a specific development model was used; 
our version identification improves on \bindiff; and
the \graphs are accurate despite the use of different packers.

\begin{table*}[t]
\centering
\small
\begin{tabular}{|l|r|rrr|rrr|rrr|r|rrr|r|}
\cline{3-16}
\multicolumn{2}{c|}{} & \multicolumn{10}{c|}{\bf Unpacking} & \multicolumn{4}{c|}{\bf Disassembly} \\
\hline
\multicolumn{1}{|c|}{\multirow{2}{*}{\bf Family}} & \multicolumn{1}{c|}{\multirow{2}{*}{\bf EXE}} & \multicolumn{3}{c|}{\bf Instructions (M)} & \multicolumn{3}{c|}{\bf Processes} & \multicolumn{3}{c|}{\bf Waves} & {\bf Time} & \multicolumn{3}{c|}{\bf $|F_u|$} & \multicolumn{1}{c|}{\bf Time} \\
\cline{3-12}\cline{13-16}
  & & {\bf Min} & {\bf Avg} & {\bf Max} & {\bf Min} & {\bf Avg} & {\bf Max} & {\bf Min} & {\bf Avg} & {\bf Max} & {\bf Avg(s)} & {\bf Min} & {\bf Avg} & {\bf Max} & {\bf Avg(s)} \\ 
\hline
Allaple   & 4,000 & 0.006 &     3.1 &   497.7 & 1 & 1.00 & 1 & 1 & 4.00 &   5 &   399 &  11&   25&  458&  59 \\ \hline 
IRCbot    &   365 &   0.1 &    54.5 &   655.9 & 1 & 1.99 & 3 & 1 & 3.87 &   5 &   460 &   5&  655&  700&  114 \\ \hline
Klez      &   750 &   6.8 &     7.8 &    11.6 & 1 & 1.00 & 1 & 1 & 4.93 &   5 &   126 &   5&  749&  810&  65 \\ \hline
Loring    &   216 &  36.2 &    49.6 &    51.9 & 2 & 2.00 & 2 & 2 & 3.99 &   4 &   361 &  84&  697&  700& 164 \\ \hline
Memery    &   113 & 172.2 &   282.6 &   298.5 & 1 & 1.00 & 1 & 2 & 2.00 &   2 & 1,483 & 127&  127&  128&  27 \\ \hline
Picsys    &   131 &   7.8 &    11.0 &    11.3 & 1 & 1.00 & 1 & 1 & 1.96 &   2 &   513 &  24&  698&  737&  109 \\ \hline
Simbot    &   214 &  29.7 &   101.0 &   202.1 & 1 & 2.00 & 3 & 2 & 4.29 &   5 &   671 &  18&   85&  160&  56 \\ \hline
Sytro     & 1,354 &   2.6 &     4.5 &     8.2 & 1 & 1.00 & 1 & 1 & 1.87 &   2 &   119 &  21&  702&  810&  33 \\ \hline
Urelas    &   206 & 0.006 &   252.8 &   924.1 & 1 & 3.61 & 7 & 1 & 9.29 & 215 &   860 &  57& 2,171& 5,951& 366 \\ \hline
VtFlooder &   444 & 0.006 & 1,379.5 & 3,439.3 & 1 & 1.00 & 1 & 1 & 2.26 & 6 & 1,927 &  10&  138& 1,058&  36 \\ \hline
\end{tabular}
\caption{Unpacking and disassembly results on packed malware.
For each family, it shows the 
executables analyzed,
instructions (in millions) traced for all malware processes,
malware processes traced;
waves for all malware processes,
unpacking runtime,
number of unpacked functions in the \dbs,
and
disassembly runtime.
}
\label{tbl:all}
\end{table*}

\section{Evaluation on Malware}
\label{sec:eval:malware}

This section evaluates our lineage approach on \numfamilies malware families. 
We first describe our dataset, then 
in Section~\ref{sec:eval:unpacking} we present the unpacking and 
disassembly results, and 
in Section~\ref{sec:eval:lineage} the lineage inference results.

\paragraph{Dataset.}
The PANDA team periodically records malware executions on a sandbox and 
makes these recordings publicly available~\cite{pandarec}. 
The malware in those recordings is unlabeled, 
i.e., its family is unknown. 
To classify the malware, we first collect the AV labels for the samples 
using VirusTotal (VT)~\cite{vt}, an online service that analyzes
files and URLs submitted by users. 
We use the AV labels as input to AVClass~\cite{avclassUrl}, 
an open-source malware labeling tool.
AVClass outputs for each sample the most likely family name and a 
confidence factor based on the agreement across AV engines~\cite{avclass}.
To compute the \graph we need a significant number of samples for the 
same family. 
Thus, we select \numfamilies malware families for which more than 100 samples 
are labeled with high confidence by AVClass: 
\allaple, \ircbot, \klez, \loring, \memery, \picsys, \simbot, \sytro, 
\urelas, and \vtflooder.
In total we evaluate on \nummalware malware samples. 
The largest number of family samples is 4,000 for \allaple and the smallest 
113 for \memery.

\subsection{Unpacking and Disassembly}
\label{sec:eval:unpacking}

Table~\ref{tbl:all} details the unpacking and disassembly results 
for the \numfamilies malware families. 
For each family, it shows: 
the number of executables analyzed; 
the number of instructions (in millions) traced for all malware processes;
the number of malware processes traced; 
the number of waves for all malware processes;
the unpacking runtime (in seconds); 
the total number of unpacked functions in the \dbs
(including short and external functions that \ida counts); and 
the disassembly runtime (in seconds).

For 6 families all samples have a single process, 
for \loring all samples have 2 processes, and 
the remaining three (i.e., \ircbot, \simbot, \urelas) 
have samples with different number of processes.
The number of waves ranges from 1 (i.e., not packed) up to 
215 for a \urelas sample.
Surprisingly, 7 families have a few samples that are not packed
(e.g., 5 out of 131 samples in \picsys), 
which may be due to the developers forgetting to pack some samples 
or to early versions not being packed. 
On average, replaying a recording takes 11 minutes, 
the slowest being \vtflooder with an average of 32 minutes per recording.
The disassembly runtime takes on average 103 seconds per sample, 
of which 82\% is due to the \bw function identification. 

\begin{table*}[t]
\centering
\small
\begin{tabular}{|l|r|r|r|r|r|r|r|r|r|r|r|r|r|r|r|}
\cline{3-16}
\multicolumn{2}{c|}{} & 
\multicolumn{2}{c|}{$|V|$} & 
\multicolumn{2}{c|}{$|E|$} & 
\multicolumn{2}{c|}{$max(|L(v_i)|)$} & 
\multicolumn{2}{c|}{$|L(v_i)=1|$} & 
\multicolumn{2}{c|}{$max(|F(v_i)|)$} & 
\multicolumn{2}{c|}{$min(|F(v_i)|)$} &  
  \multicolumn{2}{c|}{$|\bigcup F(v_i)|$} \\ 
\hline
  {\bf Family} & {\bf EXE} & {\bf spp} & {\bf raw} & {\bf spp} & {\bf raw} & {\bf spp} & {\bf raw} & {\bf spp} & {\bf raw} & {\bf spp} & {\bf raw} & {\bf spp} & {\bf raw} & {\bf spp} & {\bf raw} \\
\hline
Allaple   & 4,000 &   6 & 311 &   5 & 310 & 2,742 & 2,742 &   0 & 278 &    12 &   301 &  11 &  10 & 2,387 &  5,847 \\ \hline
IRCbot    &   365 &  11 &  12 &   4 &   5 &   338 &   338 &   8 &   9 &   510 &   545 &   2 &   2 &   700 &    980 \\ \hline
Klez      &   750 &  64 &  66 &  63 &  68 &   585 &   585 &  47 &  49 &   619 &   667 &   5 &   5 &   691 &  1,114 \\ \hline
Loring    &   216 &   2 &   3 &   1 &   2 &   215 &   127 &   1 &   1 &   510 &   545 &  60 &  61 &   521 &    726 \\ \hline
Memery    &   113 &   9 &   9 &   8 &   8 &    65 &    65 &   3 &   3 &   121 &   123 & 120 & 122 &   130 &    134 \\ \hline
Picsys    &   131 &   3 &   4 &   2 &   3 &    95 &    95 &   0 &   1 &   379 &   473 &  16 &  16 &   387 &    736 \\ \hline
Simbot    &   214 &  37 & 110 &  36 & 109 &    65 &    48 &  23 &  94 &    67 &    72 &  17 &  17 &   126 &  1,989 \\ \hline
Sytro     & 1,354 &   6 &   7 &   5 &   4 &   811 &   811 &   0 &   0 &   618 &   667 &  13 &  13 &   758 &  1,509 \\ \hline
Urelas    &   206 & 123 & 130 & 213 & 179 &    22 &    22 & 105 & 115 & 3,702 & 4,420 &  42 &  44 & 7,725 & 78,408 \\ \hline
VtFlooder &   444 &  32 &  95 &  39 &  95 &   228 &   228 &  16 &  82 &   905 &   945 &  10 &  10 & 5,324 &  9,397 \\ \hline
\end{tabular}
\caption{\Graph details. 
  For each hash, it details the number of versions, edges, 
  the maximum number of samples in a version, 
  the number of singleton versions representing a single sample, 
  the maximum and minimum number of functions in a version, and
  the total number of functions across all versions.
} 
\label{tbl:graph}
\end{table*}

\begin{figure}[t]
  \centering
\begin{tikzpicture}[->,>=stealth',shorten >=0.15pt,auto,node distance=1.50cm, semithick, every node/.style={scale=1.2, text width = 2.9em, align=center, font=\scriptsize}]

    \node[state,draw=black,text=black,inner sep=0]  (P13)                      {$13{,}66$};
  \node[state,draw=black,text=black,inner sep=0]  (P335)   [right of=P13]    {$335{,}17$};
  \node[state,draw=black,text=black,inner sep=0]  (P618_1) [right of=P335]   {$618{,}273$};
  \node[state,draw=black,text=black,inner sep=0]  (P618)   [right of=P618_1] {$618{,}811$};
  \node[state,draw=black,text=black,inner sep=0]  (P22)    [above of=P618]   {$22{,}111$};
  \node[state,draw=black,text=black,inner sep=0]  (P618_2) [right of=P618]   {$618{,}76$};

 \begin{scope}[every edge/.append={scale=0.9, font=\scriptsize}]
  \path (P13)     edge  node {13}  (P335)  
        (P335)    edge  node {215} (P618_1)
        (P618_1)  edge  node {609} (P618)  
        (P618_1)  edge  node [pos = 0.7, outer sep=-6pt] {22}  (P22)   
        (P618)    edge  node {615} (P618_2);
 \end{scope}

\end{tikzpicture}

  \caption{\Graph for \sytro built using 1,354
  samples. The node label indicates the number of functions
  in the version and the number of samples of that version.
  The edge label is the number of shared functions.
}
  \label{fig:sytro}
\end{figure}
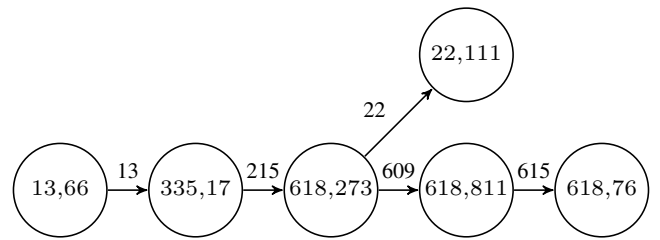

\subsection{Lineage Inference}
\label{sec:eval:lineage}

Table~\ref{tbl:graph} details the final {\graph}s 
output for the \numfamilies malware families.
The table shows the number of executables for each family and for each hash: 
the number of nodes and edges, 
the maximum number of samples in a version, the number of singleton versions representing only one sample, 
the maximum and minimum number of functions in a version, and 
the total number of functions across all versions.

We focus on the \spp hash results as Section~\ref{sec:accuracy:lineage} 
demonstrates that it outperforms the raw hash.
The largest number of \spp versions is for \urelas with 123 and the smallest 
for \loring with 2.
For 9 families, 
the number of versions using both hashes differs.
Overall, we identify 293 \spp versions across the \numfamilies families, 
compared to \nummalware input samples, 
a 26 times reduction (10x reduction for the raw hash).
Thus, our \graph is a succinct summary of family evolution, 
achieving over an order of magnitude reduction in nodes compared to 
prior approaches where each input sample is a node. 

The results show that some versions are more popular than others
in terms of samples. 
For example, 99\% of \loring and 92\% of \ircbot samples are derived 
from one version.
On the other hand, 51\% of \urelas samples correspond to singleton versions,
which may be due to a large amount of experimentation in the family, 
or to limited coverage of our input samples for some versions.
The code of a family can significantly evolve over time. 
For example \urelas versions range from 42 up to 4,420 functions. 
We also observe different code base sizes across families. 
The most complicated malware families are \urelas and \vtflooder with 
7.7K and 5.3K functions across all versions, respectively.
The simplest malware family is \simbot with a total of 126 functions 
in 216 samples. 

Of the \numfamilies malware families, 
3 have straight-line development (\allaple, \loring, \picsys). 
Figure~\ref{fig:picsys} shows the final \picsys \graph. 
\memery and \sytro have mostly straight development with only one 
node having two branches, one of them leading to a single node, 
as illustrated in the \graph for \sytro in Figure~\ref{fig:sytro}. 
\simbot has mostly straight development with 4 nodes having two branches. 
\ircbot has multiple straight lines, although 98\% of the input samples 
belong to the same line.
Finally, the \graphs of 3 families (\klez, \urelas, \vtflooder) are DAGs.
These results show the variety of development models malware 
families may use and the difficulty of predicting the development model apriori.

To conclude the results show that the \graph is a succinct summary of the 
evolution of a malware family, which reduces the number of input samples to a 
much smaller (i.e., 26 times smaller) number of versions. 
The \numfamilies malware families show varying 
development models that are captured by our lineage inference algorithm.
\section{Discussion}
\label{sec:discussion}

This section discusses some limitations of our work and 
directions for future investigation.

\paragraph{Packers that modify the original code.}
Our unpacking module handles packers where the original code is recovered at 
runtime. 
However, packers may transform the original code, so that it is no longer 
present in the packed executable. 
For example, virtualization-based packers such as Themida and VMProtect
convert assembly into bytecode interpreted by a VM. 
One approach to handle such packers is analyzing the executable code output by the interpreter. 
We leave the support of such packers as future work.

\paragraph{Evasion.}
Similar to other dynamic analysis approaches, 
our unpacking can be evaded by 
techniques that detect the presence of a VM or emulator.
We ameliorate this problem by incorporating countermeasures for specific 
anti-VM checks, 
but our countermeasures are not complete. 
Interestingly, we observe that anti-VM checks are typically packed 
themselves to avoid their presence flagging the executable as malware. 
If the anti-VM checks are unpacked simultaneously with the 
original code, by the time the malware executes the anti-VM checks 
our unpacking has already captured the original code in a \statef. 

\paragraph{Code semantics.}
Our approach enables comparing the 
functions added and removed across versions of a malware family. 
However, the analyst still needs to examine the added and removed functions 
to understand their functionality. 
In future work, we plan to provide the analyst with summaries of the 
semantics of those functions to further reduce the malware analysis effort.

\paragraph{Function identification.}
Most disassembly errors in our evaluation are due to missed functions. 
Our system currently uses \bw~\cite{byteweight} for function identification, 
but could use other tools. 
We are currently evaluating Nucleus~\cite{nucleus}, 
which should improve on \bw results, 
but have not finished integrating it into our toolchain yet.

\section{Conclusion}
\label{sec:conclusion}

We have presented a novel malware lineage approach that 
works on malware collected in the wild. 
Given a pool of malware executables from the same family,
it produces a {\em \graph} where nodes are family versions and
edges describe their descendant relationships.
We have proposed the first technique to identify different versions
of a malware family and a scalable code indexing technique for
efficiently identifying functions shared between any pair of versions.
We have evaluated the accuracy of our approach on 
13 benign programs and produced \graphs for \numfamilies malware families, 
showing that the produced \graphs are a succinct representation of 
the evolution of a malware family.

\appendix

\subsection{Complexity Analysis}
\label{sec:complexity}

Let the set of malware samples be $S= \{ s_1, \cdots, s_n \}$
and $\mathcal{F}$ be the set of functions appearing in all 
malware samples in $S$. Let $F: S \rightarrow 2^{\mathcal{F}}$
label each malware with the set of functions appearing in it
(e.g., $F(s_2) = \{ f_5 , f_8 \}$ means that functions $f_5$
and $f_8$ appear in $s_2$). Our algorithm is divided into
three phases.

\noindent
{\bf Phase I (clustering):} The goal of this phase is to find a
coarsest partition $\mathcal{P} \; = \; \{ G_1,G_2,\cdots,G_k \}$ of
$S$ such that $s_i$ and $s_j$ are in the same group iff $F(s_i) =
F(s_j)$.  Using hashing, the cost of finding this partition is $O(n)$.
We overload the function $F$, and use $F(G_i)$ to be $F(s)$
for some $s \in G_i$ (since all $s \in G(s_i)$ have the same label 
$F(s)$, picking an arbitrary $s$ does not create an issue.)

\noindent
{\bf Phase II (Lineage Tree):} This tree $T=(V,E)$ has $k$ vertices
$\{ 1, \cdots, k \}$, where vertex $i$ corresponds to group $G_i$. Let
$\mbox{inst}(s_i,s_j)$ be the number of instructions common between
$s_i$ and $s_j$. We say that $G_i \prec G_j$ iff $F(i) \subset F(j)$.
For each $j \in V$, define $\mbox{dist}(i,j)$ (for $i \in V$) to be
the $2$-tuple $\langle | F(G_i) \cup F(G_j) | , \mbox{inst}(s_i,s_j)
\rangle$. We say that $\langle n,m \rangle < \langle n',m' \rangle$
iff $n < n'$ or ($n = n' \; \wedge \; m < m'$) (i.e., we use
lexicographical ordering on the $2$-tuple). Let $\mbox{MIN}(j)
\subseteq V - \{ j \}$ be all the minimal elements according to the
metric $\mbox{dist}$ (i.e., $i \in \mbox{MIN}(j)$ iff there does no
exist a $k \in V - \{ j \}$ such that $\mbox{dist}(k,j) <
\mbox{dist}(i,j)$).  We add an edge $(i,j) \in E$ where $i$ is the
highest index in the set $\mbox{MIN}(j)$. Assuming that all the
metrics have been pre-computed, computing edges in $E$ takes at most
$O(k^2)$ time.

\noindent
{\bf Phase III (Adding cross-edges):} Next we add cross-edges to the
tree $T = (V,E)$ constructed in phase II. Le the set of cross-edges
$E_c \subseteq V \times V$. Let $F_T : V \rightarrow \mathcal{F}$ be a
such that $F_T (i)$ is $F(i)$ minus the labels of the parent of $i$
and children of $i$ (if the parent or children of $i$ don't exist,
then we do subtract anything from $F(i)$).  For each $j \in V$, let
$\mbox{Z}(j)$ be all $i \in V$ that satisfy the following conditions:
$i \not= j$ and $i$ is not the parent or child of $j$ in the lineage
tree $T$. We want to find a set of indices of minimal size $I(j)
\subseteq \mbox{Z}(j)$, such that $F_T (j) \subseteq \bigcup_{i \in
  I(j)} F(i)$ (in other words we want to find the minimal cover for
the set $F_t (j)$ using sets whose indices are in
$\mbox{Z}(j)$. Finding the minimal cover is NP-hard, so we rely on
iterative algorithm to ``approximate'' $I(j)$. We add $I(j) \times \{
j \}$ to the set of cross-edges $E_c$.  Our iterative algorithm to
pick $I(j)$ takes time $O(k)$ for each $j$. Therefore, the entire
running time of this phase is $O(k^2)$.

\end{document}